\documentclass[doublecol]{epl2}

\usepackage{amsmath}
\usepackage{amsfonts}

\usepackage{hyperref}
\usepackage{bm}
\renewcommand{\vec}{\bm}

\newcommand{\dif}{\mathrm{d}}
\newcommand{\mi}{\mathrm{i}}
\newcommand{\me}{\mathrm{e}}

\title{Bose-Einstein condensates with Raman-induced spin-orbit coupling : An overview}
\shorttitle{Bose-Einstein condensates with Raman-induced spin-orbit coupling : An overview}

\author{Giovanni Italo Martone\inst{1,2}}
\shortauthor{G. I. Martone}

\institute{
  \inst{1} CNR NANOTEC, Institute of Nanotechnology, Via Monteroni, 73100 Lecce, Italy \\
  \inst{2} INFN, Sezione di Lecce, 73100 Lecce, Italy
}

\abstract{
Since their first realization more than a decade ago spin-orbit-coupled Bose-Einstein condensates have been the subject
of intense theoretical and experimental investigations. Spin-orbit coupling deeply modifies the equilibrium properties
of the condensate, giving rise to novel configurations such as a supersolid stripe phase and a phase-separated
plane-wave state. At the level of dynamics, both the frequency and the nature of the collective modes are significantly
affected by the coupling with the spin degree of freedom. Here we review some of the most relevant advances in the field
and provide our perspective on possible future research directions.}

\begin{document}

\maketitle

\section{Introduction}
The recent years have witnessed a number of important advances in the manipulation of quantum degenerate gases.
One of the most outstanding achievements is represented by the realization of artificial gauge potentials, which enables
one to use neutral ultracold atoms to simulate the behavior of charged particles in external electric or magnetic fields
(see the reviews~\cite{Dalibard2011review,Goldman2014review} and references therein). Synthetic gauge fields can also be
engineered to act on some internal degree of freedom, such as the atomic spin; this entails a coupling between the spin
and the center-of-mass motion of each atom, which is commonly referred to as spin-orbit coupling (SOC). In condensed
matter physics SOC is known to play a crucial role in phenomena and systems such as the spin Hall effect~\cite{
Sinova2015review}, Majorana fermions~\cite{Elliott2015review} and topological insulators~\cite{Hasan2010review,Qi2011review}.
The experimental implementation of artificial SOC on cold atomic systems, first accomplished by Spielman's group
at NIST~\cite{Lin2011}, has paved the way towards the study of SOC physics in a highly controllable framework.

A major advantage of spin-orbit-coupled quantum gases is represented by the possibility of exploiting the interplay between
the single-particle dispersion (that is deeply modified by SOC) and the interparticle interaction to realize novel many-body
configurations, with no counterpart in other domains of condensed matter physics. For instance, one can engineer SOC on
a two-component Bose mixture, thus creating a spin-orbit-coupled Bose-Einstein condensate (BEC) of effective spin
$1/2$,\footnote{This degree of freedom is sometimes referred to as the pseudospin, as it does not correspond to the real
intrinsic spin of the particles.} as done in~\cite{Lin2011}. These systems present exotic ground states, including
plane-wave phases, where Bose-Einstein condensation occurs in a state with finite momentum and magnetization, and stripe
phases, exhibiting periodic modulations in the density profile. Because of these features spin-orbit-coupled BECs have
attracted a broad interest on the experimental as well as the theoretical side. Many of these aspects are summarized
in the previous reviews~\cite{Dalibard2011review,Galitski2013review,Zhou2013review,Goldman2014review,Zhai2015review,
Li2015review,Zhang2016review,Recati2021review}.

In the groundbreaking experiment~\cite{Lin2011} the authors realized a special kind of SOC, given by an equal-weighted
superposition of Rashba~\cite{Bychkov1984} and Dresselhaus~\cite{Dresselhaus1955} terms. This is often referred to as
Raman-induced spin-orbit coupling, as it results from the interaction of the BEC with a properly designed arrangement
of Raman lasers, yielding transitions between different hyperfine states of the atoms. The NIST scheme has been also
extended to degenerate Fermi gases~\cite{Wang2012,Cheuk2012} and to spin-$1$ configurations~\cite{Campbell2016}. In
addition, a similar method has led to the realization of BECs with spin-orbital-angular-momentum coupling~\cite{Chen2018b,
Zhang2019}. In all these setups the coupling between the atom motion and its spin occurs only along one direction in
space, but in more recent experiments a two and three-dimensional SOC has been created for bosons, both
with~\cite{Wu2016,Wang2021} and without a lattice~\cite{Valdes2021}, and for fermions~\cite{Meng2016}, allowing for
the observation of interesting topological features. Despite all these advances, spin-$1/2$ Bose gases with
one-dimensional SOC as realized in~\cite{Lin2011} are still a very active field of research. In this review we will
summarize some of the most relevant properties of these systems and discuss a few open questions to be addressed by
future studies.

\section{Single-particle Hamiltonian}
The experimental scheme employed by the NIST team in~\cite{Lin2011} consists of a $^{87}$Rb gas in the $F = 1$ ground electronic
manifold, with a bias magnetic field lifting the degeneracy of the levels within the manifold. A pair of Raman laser fields
imparts momentum $2 \hbar \vec{k}_R$ and energy $\hbar \Delta \omega_L$ to the system, at the same time inducing transitions
between the hyperfine levels. The wave vector $\vec{k}_R = k_R \hat{\vec{e}}_x$ lies along the $x$ direction identified by the
unit vector $\hat{\vec{e}}_x$, and the frequency detuning $\Delta \omega_L$ of the two lasers is taken close to the frequency
splitting $\omega_Z$ characterizing the transition between the $|F=1,m_F=-1\rangle$ and $|F=1,m_F=0\rangle$ levels, henceforth
called the spin-down and spin-up state, respectively. On the other hand, the separation between the $|F=1,m_F=0\rangle$
and $|F=1,m_F=+1\rangle$ levels is much larger because of an additional contribution stemming from the quadratic Zeeman effect.
Consequently, the $|F=1,m_F=+1\rangle$ state can be adiabatically eliminated and one is left with an effective spin-$1/2$
system~\cite{Lin2011}. By further performing the space and time-dependent spin rotation $\mathcal{U} = \exp \left[ \mi (2 k_R x
- \Delta\omega_L t) \sigma_z / 2 \right]$ on the single-particle Hamiltonian one can remove its dependence on the phase of the
Raman lasers~\cite{Lin2011,Martone2012}, bringing it into the time-independent form
\begin{equation}
h_{\mathrm{SO}} = \frac{\left( \hat{\vec{p}} - \hbar \vec{k}_R \sigma_z \right)^2}{2m} + \frac{\hbar\Omega_R}{2} \, \sigma_x
+ \frac{\hbar \delta_R}{2} \, \sigma_z + V_{\mathrm{ext}}(\vec{r}) \, .
\label{eq:h_SO}
\end{equation}
Here $m$ is the atom mass, $\Omega_R$ the Raman coupling fixed by the intensity of the laser beams, $\delta_R = \Delta \omega_L
- \omega_Z$ the detuning of the laser field from Raman resonance, while $\sigma_{x,y,z}$ denote the $2 \times 2$ Pauli matrices.
The operator $\hat{\vec{p}} = - \mi \hbar \nabla$ is the canonical momentum, which differs from the physical momentum $\hat{\vec{P}}
= \hat{\vec{p}} - \hbar \vec{k}_R \sigma_z$ fixing the velocity of particles due to the SOC contribution acting along $x$. Notice
that in the absence of the external trapping potential $V_{\mathrm{ext}}(\vec{r})$ Hamiltonian~\eqref{eq:h_SO} is translationally
invariant, consequently its eigenstates can be taken as plane waves multiplied by a two-component spinor. The energy spectrum is
made of two branches,
\begin{equation}
\varepsilon_\pm(\vec{p})
= \frac{\vec{p}^2}{2m} + E_R \pm \frac{\hbar}{2} \sqrt{\left( \frac{2 k_R p_x}{m} - \delta_R \right)^2 + \Omega_R^2} \, ,
\label{eq:sp_en}
\end{equation}
with $E_R = (\hbar k_R)^2 / 2m$ the recoil energy.
\begin{figure}
\onefigure{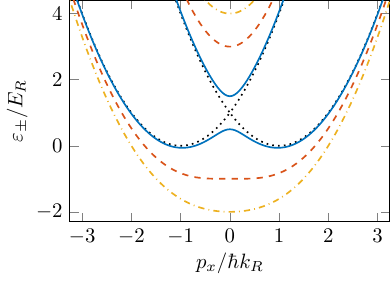}
\caption{Lower and upper branch of the single-particle spectrum~\eqref{eq:sp_en} with $\delta_R = 0$ as functions of momentum and
for $\hbar\Omega_R / E_R = 1.0$ (blue solid line), $4.0$ (red dashed line) and $6.0$ (yellow dash-dotted line). The black
dotted line shows the dispersion at zero Raman coupling $\Omega_R$.}
\label{fig:sp_spectrum}
\end{figure}
These branches as functions of momentum are plotted in Fig.~\ref{fig:sp_spectrum}
for $\delta_R = 0$ and several values of $\Omega_R$. The most relevant property of this peculiar single-particle spectrum is that for
$\hbar\Omega_R < 4 E_R$ the lower branch exhibits two degenerate minima at finite momenta $\vec{p} = \pm \hbar k_1^{(0)} \hat{\vec{e}}_x$
with $k_1^{(0)} = k_R \sqrt{1 - \left[\hbar \Omega_R / (4 E_R)\right]^2}$. Notice that $k_1^{(0)} \to 0$ when $\hbar\Omega_R$ approaches
$4 E_R$, and above this value one has a single minimum at $\vec{p} = 0$. The double-minimum structure is at the origin of novel features,
that emerge at the many-body level when the two-body contact interaction between particles is accounted for, as explained below.

\section{Many-body ground state}
The many-body ground state of a weakly interacting BEC with single-particle Hamiltonian as in Eq.~\eqref{eq:h_SO} was first
investigated in Refs.~\cite{Ho2011,Li2012a}, after earlier works had addressed the case of a pure Rashba SOC~\cite{Wang2010,Wu2011}.
These analyses make use of the Gross-Pitaevskii mean-field approach, which is justified because in dilute three-dimensional systems
with SOC along a single direction the quantum depletion does not exceed a few percent~\cite{Zheng2013,Chen2018}. In this approach
finding the ground state amounts to minimizing the total energy
\begin{equation}
E = \int_V \dif\vec{r} \left( \Psi^\dagger h_{\mathrm{SO}} \Psi + \frac{g_{dd}}{2} n^2 + \frac{g_{ss}}{2} s_z^2
+ g_{ds} n s_z \right)
\label{eq:gp_en}
\end{equation}
with respect to the two-component condensate order parameter $\Psi$. Here $V$ is the volume occupied by the system, while
$n = \Psi^\dagger \Psi$ and $s_z = \Psi^\dagger \sigma_z \Psi$ are the total and spin density, respectively. The former is
normalized to the particle number, $\int_V \dif\vec{r} \, n(\vec{r}) = N$. The interaction terms feature the density-density
[$g_{dd} = (g_{\uparrow\uparrow} + g_{\downarrow\downarrow} + 2 g_{\uparrow\downarrow}) / 4$], spin-spin
[$g_{ss} = (g_{\uparrow\uparrow} + g_{\downarrow\downarrow} - 2 g_{\uparrow\downarrow}) / 4$], and density-spin
[$g_{ds} = (g_{\uparrow\uparrow} - g_{\downarrow\downarrow}) / 4$] coupling constants, where $g_{\sigma\sigma'} =
4 \pi \hbar^2 a_{\sigma\sigma'} / m$ ($\sigma,\sigma' = \uparrow,\downarrow$) are the couplings in the various spin
channels and $a_{\sigma\sigma'}$ the respective scattering lengths. Unless otherwise specified, we henceforth assume
equal intraspecies couplings, i.e., $g_{ds} = 0$, and vanishing $\delta_R$.

\begin{figure}
\onefigure{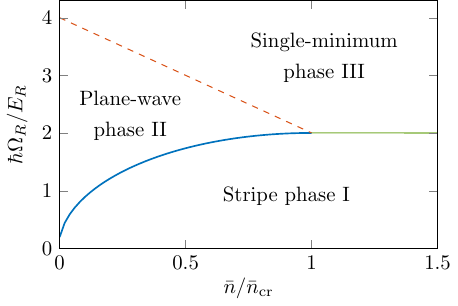}
\caption{Phase diagram of a spin-orbit-coupled Bose-Einstein condensate as a function of the average density
$\bar{n}$ and the Raman coupling $\Omega_R$. Solid (dashed) lines correspond to transitions of first (second) order.
Adapted from~\cite{Martone2012}.}
\label{fig:phase_diag}
\end{figure}

At equilibrium the order parameter evolves in time as $\Psi(\vec{r},t) = \me^{- \mi \mu t / \hbar} \Psi_0(\vec{r})$, with
$\mu$ the chemical potential. For infinite systems ($V_{\mathrm{ext}} = 0$) one can formulate an Ansatz for $\Psi_0$ as a
superposition of two counterpropagating plane waves of the form
\begin{equation}
\Psi_0(\vec{r}) = \sqrt{\bar{n}} \left[ C_+
\begin{pmatrix}
\cos\vartheta \\
- \sin\vartheta
\end{pmatrix}
\me^{\mi k_1 x}
+ C_-
\begin{pmatrix}
- \sin\vartheta \\
\cos\vartheta
\end{pmatrix}
\me^{- \mi k_1 x}
\right] .
\label{eq:ansatz}
\end{equation}
Here $\bar{n} = N/V$ is the average density, $C_\pm$ are complex weights normalized as $|C_+|^2 + |C_-|^2 = 1$, and
$0 \leq \vartheta \leq \pi/4$ quantifies the spin polarization of the two momentum components of the wave function.
The wave vector $k_1$ is related to the polarization as $k_1 = k_R \cos 2\vartheta$, which follows from energy
minimization (see below). Notice that Eq.~\eqref{eq:ansatz} corresponds to the exact ground state of the system in
the absence of interaction, with arbitrary $C_\pm$ and $k_1$ coinciding with its single-particle value $k_1^{(0)}$.
It is thus natural to expect that this Ansatz can accurately describe interacting systems at sufficiently low densities,
provided that the parameters $C_\pm$, $\vartheta$ and $k_1$ are chosen so as to minimize the energy~\eqref{eq:gp_en}.
A major difference with respect to the ideal gas case is that even an infinitesimally weak interaction can lift
the degeneracy with respect to the relative populations $|C_\pm|^2$ of the two momentum states. Indeed, the main result
of the energy minimization procedure is that the phase diagram of a spin-orbit-coupled BEC contains three distinct
quantum phases~\cite{Ho2011,Li2012a}, that can be accessed by tuning, e.g., the Raman coupling $\Omega_R$ or the
average density $\bar{n}$, see Fig.~\ref{fig:phase_diag}.
The phases are the following:
\begin{enumerate}
\renewcommand{\labelenumi}{\Roman{enumi}.}
\item Stripe phase, where the condensate wave function~\eqref{eq:ansatz} features equal occupation of the two states
with opposite momenta, $|C_+|^2 = |C_-|^2 = 1 / 2$, resulting in a vanishing spin polarization $\langle \sigma_z \rangle
= \int_V \dif\vec{r} \, s_z(\vec{r})$. By interfering, the two counterpropagating plane waves give rise to modulations
(stripes) in the density profile oscillating at wavelength $\pi / k_1$, where $k_1 = k_R \sqrt{1 - [\hbar \Omega_R
/ (4 E_R + g_{dd} \bar{n})]^2}$.
\item Plane-wave (PW) phase, where the atoms condense into just one of the two momentum states in the Ansatz~\eqref{eq:ansatz}.
The two possible realizations of this configuration correspond to a vanishing value of either $C_-$ or $C_+$. They have
uniform density and opposite momenta $\vec{p} = \pm \hbar k_1 \hat{\vec{e}}_x$ and polarizations $\langle \sigma_z \rangle
= \pm N k_1 / k_R$, with $k_1 = k_R \sqrt{1 - [\hbar \Omega_R / (4 E_R - 2 g_{ss} \bar{n})]^2}$.
\item Single-minimum (SM) phase, characterized by a uniform density and a vanishing condensation momentum $k_1$ and spin
polarization $\langle \sigma_z \rangle$.
\end{enumerate}
The stripe phase, which appears only if $g_{ss} > 0$, is of great interest because it spontaneously breaks translation
invariance and thus exhibits supersolid-like properties, as discussed in more details below. This configuration is
sometimes referred to as the mixed phase because the two momentum components in the wave function~\eqref{eq:ansatz}
form a coherent superposition and thus fully overlap in real space.

When the Raman coupling $\Omega_R$ exceeds a critical value given, at low densities, by $\hbar\Omega_{\mathrm{cr1}}
= 4 E_R \sqrt{2 g_{ss} / (g_{dd} + 2 g_{ss})}$ \cite{Ho2011,Li2012a}, the system undergoes a first-order phase
transition to the PW phase. This state is twofold degenerate as it breaks a $\mathbb{Z}_2$ symmetry of the energy
functional~\eqref{eq:gp_en} (namely, the one represented by the operator $\sigma_x \mathcal{P}$ that simultaneously
performs a spin flip and a parity transformation $\mathcal{P}$ on the wave function $\Psi$). Interestingly, in experiments
where the Raman coupling is adiabatically ramped up starting from the stripe phase one observes a phase separation effect
above $\Omega_{\mathrm{cr1}}$, with the two plane-wave components of the order parameter~\eqref{eq:ansatz} 
occupying different regions in space~\cite{Lin2011,Putra2020}, while keeping the system globally unpolarized. In
this case density fringes can occur only in the domain wall, where the components still overlap, as shown
in~\cite{Putra2020}. Ferromagnetic domains with opposite momenta can also form in BECs in shaken optical
lattices~\cite{Parker2013}, whose energy dispersion can exhibit a double-minimum structure similar to that of
Fig.~\ref{fig:sp_spectrum}.

By further increasing the Raman coupling the plane-wave momentum and polarization decrease (in absolute value) and vanish
at the critical value $\hbar\Omega_{\mathrm{cr2}} = 4 E_R - 2 g_{ss} \bar{n}$, where a second-order phase transition occurs
and the system enters the SM phase. This transition was observed in~\cite{Lin2011}, by measuring the momentum
distribution of the BEC as a function of $\Omega_R$, and in~\cite{Hamner2014}, where the behavior of the spin polarization
was studied, and a connection with the superradiant-to-normal phase transition in the Dicke model was pointed out. Another
signature of the transition is represented by the divergence of the magnetic susceptibility, i.e., the response $\chi_M =
\lim_{h \to 0} \dif \langle \sigma_z \rangle / \dif h$ of the system's spin polarization to an external perturbation
$- h \sigma_z$ added to the single-particle Hamiltonian~\eqref{eq:h_SO}. The magnetic susceptibility, first computed
in~\cite{Li2012b}, is given by $\chi_M^{(\mathrm{II})} = 2 \Omega_R^2 / [\hbar\Omega_{\mathrm{cr2}}
\left(\Omega_{\mathrm{cr2}}^2 - \Omega_R^2\right)]$ in the PW phase and by $\chi_M^{(\mathrm{III})} =
2 / [\hbar\left(\Omega_R - \Omega_{\mathrm{cr2}}\right)]$ in the SM phase. This prediction was found in very good agreement
with the value measured in the experiment~\cite{Zhang2012} through the study of the center-of-mass oscillation, see below.
For $g_{ss} = 0$ the magnetic susceptibility is related to the effective mass $1/m^* = \partial^2 \varepsilon_- /
\partial p_x^2|_{\vec{p} = \pm \hbar k_1 \hat{\vec{e}}_x}$, characterizing the curvature of the single-particle
spectrum~\eqref{eq:sp_en} close to the minima, as $m^* / m = 1 + 2 E_R \chi_M$.

An intriguing property of the spin-orbit Hamiltonian~\eqref{eq:h_SO} is that, despite being invariant under translations,
it lacks Galilean invariance because it does not commute with the $x$ component of the physical momentum $\hat{\vec{P}}$. 
This has important consequences on the superfluid behavior of the system. As pointed out in~\cite{Zhang2016,Chen2018},
the superfluid density of a spin-orbit-coupled BEC is an anisotropic diagonal tensor, whose transverse components
$\rho_s^y = \rho_s^z$ match the total mass density $\bar{\rho} = m \bar{n}$, while the longitudinal component $\rho_s^x$
is smaller than $\bar{\rho}$ even at zero temperature. In the PW and SM phases one has $\rho_s^x =
\bar{\rho} / (1 + 2 E_R \chi_M)$~\cite{Zhang2016,Chen2018}, signifying that the superfluid density is strongly suppressed
as one approaches the PW--to--SM phase transition where $\chi_M$ diverges. This peculiar effect could be detected
experimentally by measuring the square of the ratio of the sound velocities along and perpendicular to the $x$
direction~\cite{Martone2021}. It is worth mentioning that the rotational properties are affected in a similar way:
as shown in~\cite{Stringari2017,Qu2018}, in isotropic harmonic traps the momentum of inertia, which vanishes for a
standard BEC, is finite in the presence of SOC, and takes its rigid value at the PW--to--SM phase transition.

Although the above results hold at zero temperature, thermal effects have been studied both theoretically~\cite{Yu2014}
and experimentally~\cite{Ji2014}. For example, it has been found that finite temperature reduces the critical value
$\Omega_{\mathrm{cr1}}$, thus favoring the PW over the stripe phase~\cite{Yu2014,Ji2014}. Back to zero temperature,
in~\cite{Li2012a} it was found that the phase diagram exhibits a nontrivial density dependence. In particular, by increasing
$\bar{n}$ up to a critical value $\bar{n}_{\mathrm{cr}}$ one can achieve a tricritical point where the stripe, PW and SM
phases coexist, as shown in Fig.~\ref{fig:phase_diag}; for $\bar{n} > \bar{n}_{\mathrm{cr}}$ the PW phase disappears, and
by increasing the Raman coupling one can induce a direct first-order transition from the stripe to the SM phase. The
existence of the tricritical point has been later confirmed by Monte Carlo calculations~\cite{Sanchez2020}, which
interestingly reveal a strong enhancement of the stability of the stripe phase due to interatomic correlations, yielding
a significant decrease of the value of $\bar{n}_{\mathrm{cr}}$ as compared to the mean-field prediction.

\section{Collective modes in non-supersolid phases}
As a result of their rich phase diagram, spin-orbit-coupled Bose-Einstein condensates exhibit a nontrivial dynamical behavior
in both infinite and trapped setups. The time evolution of the condensate order parameter $\Psi$ is governed by the time-dependent
Gross-Pitaevskii equation~\cite{Pitaevskii_Stringari_book}
\begin{equation}
\mi \hbar \partial_t \Psi =
\left[ h_\mathrm{SO} + g_{dd} \left(\Psi^\dagger \Psi\right)
+ g_{ss} \left(\Psi^\dagger \sigma_z \Psi\right) \sigma_z \right] \Psi \, .
\label{eq:td_gp_eq}
\end{equation}

Novel features already emerge at the level of the Bogoliubov theory describing the dynamics of small fluctuations on top of a
given equilibrium configuration. For this purpose one writes the order parameter as $\Psi(\vec{r},t) = \me^{- \mi \mu t / \hbar}
\left[ \Psi_0(\vec{r}) + \delta \Psi(\vec{r},t) \right]$, with $\Psi_0$ the equilibrium stationary wave function introduced
earlier. The fluctuation $\delta \Psi$ obeys the linearized version of Eq.~\eqref{eq:td_gp_eq}, whose solutions are the Bogoliubov
modes of the condensate~\cite{Pitaevskii_Stringari_book}.

\begin{figure}
\onefigure{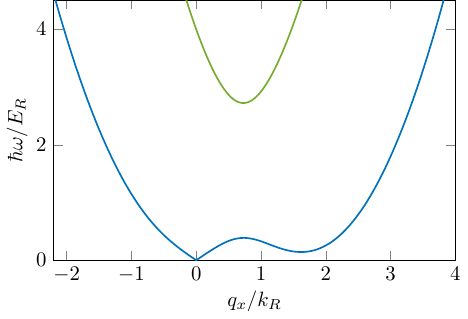}
\caption{Lower (blue) and upper (green) branch of the excitation spectrum on top of the PW phase with negative momentum.
Notice the phonon and roton structures in the lower branch. Adapted from~\cite{Martone2012}.}
\label{fig:pw_spectrum}
\end{figure}

In infinite systems, the Bogoliubov modes in the PW and SM phases are classified in terms of their momentum
$\hbar\vec{q}$ relative to the condensate; the corresponding frequency spectrum was computed in~\cite{Zheng2012,Martone2012,
Zheng2013} and measured in~\cite{Khamehchi2014,Ji2015}. Like the single-particle dispersion~\eqref{eq:sp_en}, it features
a lower gapless and an upper gapped branch (see Fig.~\ref{fig:pw_spectrum}).
At long wavelengths the gapless branch has a linear behavior, $\omega_-(\vec{q}) \simeq c q$, with anisotropic sound velocity $c$
depending on the angle between $\vec{q}$ and the $x$ axis. The velocities of excitations propagating along the positive and negative
$x$ direction fulfill the relation $m c_{x,+} c_{x,-} = \kappa^{-1} / (1 + 2 E_R \chi_M)$, with $\kappa^{-1} = \bar{n} (\partial \mu
/ \partial \bar{n})$ the inverse compressibility~\cite{Martone2012}, hence they are strongly reduced close to the PW--to--SM
transition where the magnetic susceptibility diverges; in the PW phase one additionally has $c_{x,+} \neq c_{x,-}$ if $g_{ss}
\neq 0$. The sound velocity in the directions perpendicular to $x$ is instead given by $m c_\perp^2 = \kappa^{-1}$ as in standard
BECs.

The behavior of collective modes in the presence of a harmonic trap $V_{\mathrm{ext}}(\vec{r}) = m(\omega_x^2 x^2 + \omega_y^2 y^2
+ \omega_z^2 z^2) / 2$ with frequencies $\omega_{x,y,z}$ is affected in a similar way. For example, the dipole oscillation is coupled
to the spin variable since the commutator $[h_{\mathrm{SO}},x] = - \mi \hbar (p_x - \hbar k_R \sigma_z) / m$, involving the $x$
component of the kinetic momentum $\hat{\vec{P}}$, explicitly depends on SOC. Using a sum-rule approach one finds the expression
$\omega_D = \omega_x / (1 + 2 E_R \chi_M)$ for the dipole oscillation frequency~\cite{Li2012b}, in good agreement with the experimentally
determined values of~\cite{Zhang2012}. Analogous results hold for the breathing mode~\cite{Chen2017,Geier2021}. In the $g_{ss} = 0$
case one can prove through an hydrodynamic approach that all the solutions for the collective modes derived in~\cite{Stringari1996}
for standard BECs also hold in the presence of SOC, provided that one replaces $\omega_x$ with
$\omega_x \sqrt{m / m^*}$~\cite{Martone2012}.

Another remarkable feature of the Bogoliubov spectrum of Fig.~\ref{fig:pw_spectrum} is the occurrence of a roton minimum at wave
vector $\vec{q} \sim - 2 k_1 \hat{\vec{e}}_x$ ($+ 2 k_1 \hat{\vec{e}}_x$) for the PW state with momentum $+ \hbar k_1 \hat{\vec{e}}_x$
($- \hbar k_1 \hat{\vec{e}}_x$). This minimum, like the similar one exhibited by BECs in shaken lattices~\cite{Ha2015}, is
related to the presence of an empty degenerate state that can be populated at low energy cost. The roton gap is an interaction effect,
as it vanishes in ideal gases, and softens as one approaches the transition to the stripe phase, revealing the tendency of the system
towards crystallization~\cite{Martone2012,Khamehchi2014,Ji2015}.

\section{Condensates in optical lattices}
Condensates with SOC have also been realized in the presence of periodic potentials. In a first experiment carried out by Engels'
group at WSU~\cite{Hamner2015} a BEC was loaded in a one-dimensional lattice moving along $x$ to probe its dispersion relation.
By varying the lattice velocity one observes dynamical instabilities close to the avoided crossings of the dispersion, with
different behavior along the positive and negative $x$ direction because of the lack of Galilean invariance. Subsequent studies
have considered condensates subject to a weak static potential $V_{\mathrm{ext}}(\vec{r}) = V_L \sin^2(k_L x)$, with $k_L$ and
$V_L$ the lattice wave vector and strength, respectively. The mean-field ground state is still well described by the
Ansatz~\eqref{eq:ansatz}, with $\cos\vartheta$ and $\sin\vartheta$ replaced by two $2\pi / k_L$-periodic functions~\cite{Chen2016,
Martone2016b,Martone2017}. The three phases appearing at small $V_L$ are strictly related to those found in the absence of the
lattice: a mixed phase represented by a superposition of two Bloch waves with opposite quasimomenta, a polarized Bloch-wave phase 
with finite quasimomentum, and a zero-quasimomentum phase. By increasing $V_L$ one can achieve a phase transition to an unpolarized
configuration with large contrast and quasimomentum lying at the edge of the first Brillouin zone, i.e. equal to $k_L$. This state,
which can be thought of as an artificially induced stripe phase, was detected in~\cite{Bersano2019}. The transition to the
artificial stripe phase is accompanied by a strong suppression of the sound velocity~\cite{Martone2017} and the occurrence of a
gapped pseudo-Goldstone mode in the excitation spectrum~\cite{Li2021}, which could be the object of future experimental investigations.
In addition, it would be interesting to explore the regime of large lattice strength, where Mott insulator phases appear for
sufficiently strong interactions, leading to a complex phase diagram~\cite{Yamamoto2017}.

\section{Supersolid stripe phase}
As mentioned earlier, the phase diagram of a spin--orbit-coupled BEC includes a stripe phase where superfluidity appears in
conjunction with a crystalline structure, that manifests in the form of a periodic density modulation along the direction of the
SOC~\cite{Ho2011,Li2012a}. Consequently, this configuration features the spontaneous breaking of both global phase symmetry
and translation invariance, which is the defining property of supersolidity~\cite{Thouless1969,Andreev1969,Leggett1970,
Kirzhnits1971}. This phenomenon has long been sought in solid helium~\cite{Balibar2010review,Boninsegni2012review},
but without success; only recently it has been observed in BECs inside optical resonators~\cite{Leonard2017} and with
SOC~\cite{Li2017,Putra2020}, as detailed below. These earlier successes were subsequently followed by the realization
of supersolid configurations in dipolar Bose gases~\cite{Tanzi2019a,Boettcher2019,Chomaz2019}, making supersolidity one of the
most relevant research topics in the field of ultracold gases.

The mechanism leading to the appearance of a supersolid phase in spin--orbit-coupled BECs differs from that of single-component
systems, such as dipolar gases or BECs inside optical resonators, as it involves the spin degree of freedom. It is based on the
interplay between the double-minimum structure of the single-particle spectrum (Fig.~\ref{fig:sp_spectrum}) and the two-body
interaction. The former allows for the possibility of condensing in a ground state of the kind~\eqref{eq:ansatz}, having two
components with opposite momenta and polarizations; on the other hand, for $g_{ss} > 0$ the interaction favors the simultaneous
occupation of both plane-wave components of this coherent superposition, yielding an unpolarized configuration that minimizes the
spin-dependent term of the energy~\eqref{eq:gp_en}. This term actually competes with the density-dependent one, favoring states
with uniform density, which becomes more and more important as $\Omega_R$, and thus the contrast of the stripes, increases. This
competition eventually results in the first-order transition to the polarized PW phase, as discussed above. In ~\cite{Lin2011}
this transition was observed at a value $\hbar\Omega_{\mathrm{cr1}} = 0.19 \, \! E_R$ in excellent agreement with the theoretical
prediction of~\cite{Ho2011,Li2012a}. The smallness of this critical value follows from that of the ratio $g_{ss} / g_{dd}$, which
is about $10^{-3}$ for the two atomic states of ${}^{87}$Rb employed in the NIST experiment, testifying their low degree of
miscibility. In turn this implies that the maximum achievable contrast of the fringes, which is proportional to $\Omega_{\mathrm{cr1}}
/ E_R$, is very weak in the conditions of~\cite{Lin2011}, and additionally the stripe phase is very fragile against fluctuations of
external magnetic fields. For these reasons the density modulations could not be observed directly in~\cite{Lin2011}.

Due to its connection with supersolidity, the detection of the stripe phase has been an open problem for a long time. In order to
increase miscibility one can employ atomic species with tunable $g_{ss}$. Alternatively, one can consider a quasi-two-dimensional
setup with reduced spatial overlap of the two spin components along the strongly confined direction. This can be achieved through a
properly chosen spin-dependent trapping potential~\cite{Martone2014a,Martone2014b,deHond2022} or using pseudospin orbital states in a
superlattice~\cite{Li2016}. The latter method was employed in the work~\cite{Li2017} by Ketterle's group at MIT, which reported on
the first observation of the density fringes in the stripe phase through Bragg measurements; in the same experiment it was also
verified that an higher miscibility enhances the stability of the stripe phase against magnetic field fluctuations~\cite{Martone2014b}.
A further strategy to magnify the stripe amplitude consists in rapidly ramping $\Omega_R$ up to some very large value, as done
in the subsequent experiment~\cite{Putra2020}, which additionally demonstrated the spatial coherence of spin-orbit-coupled BECs.

\begin{figure}
\onefigure{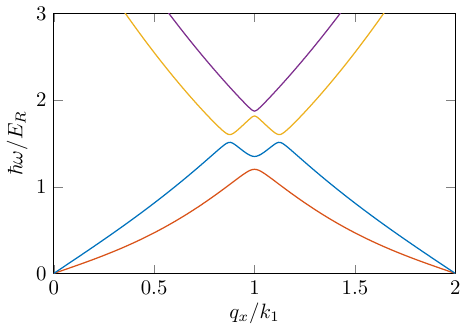}
\caption{Lower-frequency part of the excitation spectrum in the stripe phase, revealing the phonon-like behavior exhibited by the two
lowest bands at small $q_x$. Adapted from~\cite{Martone2021} under the terms of the \href{http://creativecommons.org/licenses/by/4.0/}
{Creative Commons Attribution 4.0 International License}.}
\label{fig:stripe}
\end{figure}

The stripe phase exhibits intriguing properties also at the level of its dynamic behavior. The Bogoliubov spectrum in infinite systems
as a function of the excitation quasimomentum $\hbar\vec{q}$ was computed in Refs.~\cite{Li2013,Martone2021} and is shown in
Fig.~\ref{fig:stripe}.
It features a band structure with two gapless bands, having linear dispersion at small $q$ and vanishing frequency at the edge of the
Brillouin zone $q_x = 2 k_1$. The existence of two distinct Goldstone modes, of predominantly density and spin nature, is a major
difference with respect to the non-supersolid phases (where a single gapless branch of hybridized density and spin nature is present,
see above). Their connection with the two spontaneously broken symmetries in the stripe phase is well illustrated by their $q \to 0$
behavior~\cite{Martone2021}. On the one hand, the density mode corresponds in this limit to a global phase rotation, hence it is
associated with superfluidity; on the other hand, the spin mode describes an infinitesimal translation of the stripes and has thus a
crystal-like character. In both infinite and trapped configurations the crystal Goldstone mode can be excited by suddenly releasing a
small perturbation proportional to $\sigma_z$, inducing a translational motion of the stripe at constant velocity~\cite{Geier2021,
Geier2022}. In addition, the release of a perturbation of the kind $\sigma_z \sin(q_x x)$ [$\sigma_z \sin(q_y y)$] in infinite systems
induces oscillations of the fringe spacing (orientation), and the same happens in trap with an $x \sigma_z$ ($y \sigma_z$) perturbation,
meaning that the stripes form a nonrigid and fully dynamic crystal~\cite{Geier2022}. The band structure of the spectrum of
Fig.~\ref{fig:stripe} also affects superfluidity. Indeed, despite having a finite superfluid fraction, which in the stripe phase is
smaller than unity because of the lack of both translation and Galilean invariance~\cite{Chen2018,Martone2021}, an impurity moving along
the direction perpendicular to the stripes experiences a drag force at arbitrarily small velocities~\cite{Martone2018}. Most of these
predicted features are within reach of current experiments and might become the object of future analyses.

\section{Conclusion}
In this perspective we have provided an overview of the peculiar properties of Bose gases with Raman-induced spin-orbit coupling. 
Bose condensation in these systems can occur in a plane-wave state with zero or finite momentum, as well as in a supersolid-like
stripe configuration. Spin-orbit coupling also affects the excitation spectrum and the nature of the collective modes of the
condensate, with well-pronounced effects close to the phase transitions, both in infinite and trapped setups.

Despite many efforts to understand the static and dynamic properties of these systems, several problems are still open for
investigation. Among the most promising research directions we mention the exploration of the tricritical point in the phase
diagram, the dynamics in an optical lattice, and especially the Bogoliubov modes in the stripe phase, whose observation would
advance our understanding of supersolidity in systems with a nontrivial spin degree of freedom.

\acknowledgments
The author wishes to dedicate this article to the memory of Prof. Lev P. Pitaevskii. Many useful discussions and
collaborations with K. T. Geier, P. Hauke, W. Ketterle, Y. Li, S. Stringari and D. Trypogeorgos are acknowledged.
This work was supported by the Italian Ministry of University and Research (MUR) through the PRIN project INPhoPOL
(grant 2017P9FJBS) and the PNRR MUR project PE0000023 - NQSTI.

\end{document}